\documentclass[12pt,twoside]{article}
\usepackage{fancyheadings}
\usepackage{epsfig}
\usepackage{color}
\usepackage{wtopi}

\newcommand{\rvec}[1]{\mathbf{#1}}

\newcommand{\dprime}{\prime\prime}

\newcommand{\Go}{\mathbf{G_0}}

\newcommand{\G}{\mathbf{G}}

\newcommand{\V}{\mathbf{V}}

\renewcommand{\Pi}{\mathbf{P_1}}

\newcommand{\Dw}{\tilde{D}(\omega)}

\newcommand{\fint}{\int_{-\infty}^{\infty}}

\newcommand{\zp}{z^\prime}
\newcommand{\zpp}{z^{\dprime}}
\newcommand{\zo}{z_0}
\newcommand{\w}{\omega}
\newcommand{\GVG}{(\Go\V\Go)_{RS}}
\newcommand{\GVGVG}{(\Go\V\Go\V\G)_{RS}}

\newcommand{\beq}{\begin{equation}}
\newcommand{\eeq}{\end{equation}}
\newcommand{\bea}{\begin{eqnarray}}
\newcommand{\eea}{\end{eqnarray}}

\begin{document}
\pagestyle{fancyplain}
\lhead[\thepage\hspace{10mm}\sl R.~B.~Schlottmann]{}
\rhead[]{{\sl Iterative inverse propagation}\hspace{10mm}\thepage}
\cfoot{}

\title{Direct waveform inversion by iterative inverse propagation}

\author{R.~Brian Schlottmann}

\maketitle

\section*{Summary}
Seismic waves are the most sensitive probe of the Earth's interior we have.  With the dense data sets available in exploration, images of 
subsurface structures can be obtained through processes such as migration.  Unfortunately, relating these surface recordings to actual Earth
properties is non-trivial.  Tomographic techniques use only a small amount of the information contained in the full seismogram and result in 
relatively low resolution images.  Other methods use a larger amount of the seismogram but are based on either linearization of the problem, 
an expensive statistical search over a limited range of models, or both.  We present the development of a new approach to full waveform 
inversion, i.e., inversion which uses the complete seismogram.  This new method, which falls under the general category of inverse scattering, 
is based on a highly non-linear Fredholm integral equation relating the Earth structure to itself and to the recorded seismograms.  An 
iterative solution to this equation is proposed.  The resulting algorithm is numerically intensive but is deterministic, i.e., random searches 
of model space are not required and no misfit function is needed.  Impressive numerical results in 1D are shown for several test cases.

\section{Introduction}

Seismic waves provide the best and most direct probe available of the properties of the Earth's interior.  In exploration seismology
artificially generated waves recorded at the Earth's surface are commonly used to image material discontinuities in the subsurface 
(e.g.~via migration).  However, a more elusive goal is inversion, the determination of Earth properties such as wave velocity and density 
from seismic data.  Tomographic inversions, which use only the traveltime information of isolated arrivals, are frequently used to obtain
smooth, low resolution estimates of the wave velocity generally for use in imaging algorithms.  Inversion methods which are designed to 
make use of all the information in the data, i.e., not only the arrival time and amplitude of pulses but the details of their shape as well, 
fall under the category of ``waveform inversion''.  The additional information contained in the waveforms promises to improve the accuracy
and resolution of Earth structure models.  

Although elegant in theory, waveform inversion has not enjoyed much use partly because it is computationally expensive---more than an 
order of magnitude more expensive than migration---and partly because it requires expert user intervention to make it work.  However, 
in more recent years, faster computers, better inversion strategies (e.g. Pratt, 1999a, 1999b; Sirgue and Pratt, 2004; Yokota and 
Matsushima, 2004), and a greater demand for accurate information on subsurface structure have made this approach more and more appealing.

Most of the work done to date in waveform inversion falls into one or both of two broad categories: A) linearized methods and B) model-searching
methods.  Model-searching methods (see e.g. Sen and Stoffa, 1995) involve three main components: 1) the definition of a misfit or error 
function that measures the difference between seismic waveform data and synthetic seismograms for a proposed Earth model;  2) a pre-defined 
set of adjustable model parameters and parameter ranges; and 3) some rule or set of rules for searching the defined parameter space.  The 
general idea is to find the combination of model parameters that minimizes the error.  This sort of approach can work well if the number 
of possible parameter combinations to be examined is small.  Otherwise, given that the generation of synthetic seismograms is usually quite 
expensive computationally in three spatial dimensions, searching a large-dimensional model space is impractical at this time.

Linearized methods are, on the other hand, quite cheap computationally.  A mainstay of this approach is to assume that the true, unknown Earth
structure differs only slightly from some initial reference model.  Under this assumption, one can assume that the true wavefield can be
approximated by a combination of the wavefield in the reference model and a singly-scattered wavefield.  This single-scattering assumption
is known as the first-order Born approximation.  Mathematically, this is written as
\beq
\psi(\rvec{x},\w)\approx\psi_0(\rvec{x},\w)+\w^2\int d^3\rvec{x}^\prime G_0(\rvec{x},\w;\rvec{x}^\prime)
V(\rvec{x}^\prime)\psi_0(\rvec{x}^\prime,\w),\label{Born1}
\eeq
where $\psi$ and $\psi_0$ are the actual and reference wavefields, respectively; $G_0$ is the Green's function 
of the reference medium; $V$ is the difference between the reference and actual media (and is frequently known as
the ``scattering potential''); and $\w$ is the angular frequency.  Because this
approximation is linear in $V$, one can construct algorithms to extract $V$ and, hence, an approximation to the true model from measurements
of $\psi$ on the Earth's surface.  Of course, there are many practical considerations which can make this process difficult to perform, not
the least of which is insufficient data.  However, even from a theoretical perspective, there are problems.  Specifically, if the true
structure differs more than a little from the reference model---the usual case---the data can contain arrivals, most notably multiples, that 
cannot be predicted from the single scattering approximation. 

One approach to waveform inversion that does not fall into either the linearized or model-searching categories is what is known as
inverse scattering.  The name, ``inverse scattering'', comes from the knowledge that wavefields can be viewed as resulting from multiple 
scattering of wave energy within the Earth.  The object is to find some way to invert these multiple 
scatterings to retrieve the Earth structure.  The principal work done in this area uses an approach called the ``inverse scattering series''.

Originally developed by Jost and Kohn (1952) in quantum physics and later by Moses (1956), the properties of the inverse scattering series have
been studied by Prosser (1980), among others.  Razavy (1975) was first to apply this method to the seismic inverse problem. In recent years,
it has been most extensively explored by Weglein and collaborators (e.g. Weglein et al., 2003).  The main idea
of this approach is first to write the wavefield as measured on the Earth's surface as a Born series, the multiple-scattering generalization
of eq.~(\ref{Born1}).  In a compact symbolic notation, this is written as
\beq
(\psi)_S=(\psi_0)_S+(G_0V\psi_0)_S+(G_0VG_0V\psi_0)_S+\ldots,
\eeq
where the subscript indicates that we evaluate the wavefield only where we can actually measure it.  In principle all orders of scattering
off the unknown structure $V$ are included in this sum.  If we also assume that the quantity $V$ can be written as an infinite series,
\beq
V=V_1+V_2+V_3+\dots
\eeq
where $V_n$ is $n$-th order in the recorded data, we can insert this series into the Born series and collect terms of common order in the data.
The result is an infinite set of equations for the $V_n$:
\bea
(\psi-\psi_0)_S&=&(G_0V_1\psi_0)_S\nonumber\\
0&=&(G_0V_2\psi_0)_S+(G_0V_1G_0V_1\psi_0)_S\nonumber\\
&\vdots&
\eea
These equations are meant to be solved in a cascade fashion, solving first for $V_1$ then using that result to obtain $V_2$ and so forth.

By studying these equations numerically in one spatial dimension, Carvalho (1992) has found that this approach does not appear to be convergent
for an arbitrary contrast between the reference and the actual medium. Following his result, Weglein and collaborators have studied the
sub-series approach, in which various terms in the full inverse scattering series of common form are combined into separate sub-series, each
of which is postulated to perform a different task corresponding to a conventional processing step (Weglein et al., 2003).
Using the sub-series approach, they have been able to achieve some promising results with no obvious problems with divergence.

We have derived an alternative approach to the problem of waveform inversion---one which also falls into the category of inverse scattering.  
We introduce this new development in the next section. 

\section{Theory}

\subsection{Derivation}

To introduce and test the concepts of our approach, we present the theory in one spatial dimension.
The data will consist of a single trace recorded at $z=0$ for a source also located at $z=0$.  The source-time function
will be a $\delta$-function, making this single trace the complete impulse response for the chosen source/receiver pair.
The unknown structure we will wish to recover is assumed to be located ``beneath'' the source/receiver, along the half-line $z>0$.
Further constraints we impose on the physics of the problem are that we have purely acoustic propagation, fully variable velocity,
no attenuation, and constant density.  This last constraint is imposed not only for simplicity but for the reason that it is not 
possible to reconstruct both velocity \emph{and} density from a single trace in 1D.

We take as our reference, or background, a constant-velocity medium with wave propagation velocity $c_0$ and associated Green's 
function $G_0$.  Let $c(z)$ and $G$ be the wave velocity and Green's function of the true medium, respectively.  We define the scattering
potential as
\beq
V(z)=\frac{c^2_0}{c^2(z)}-1.
\eeq
The basis of our derivation will be what is known as the Lippmann-Schwinger equation, the integral equation equivalent to
the more frequently seen differential equation for acoustic wave propagation:
\beq
G(z,\w;\zo)=G_0(z,\w;\zo)+\frac{\w^2}{c^2_0}\fint d\zp G_0(z,\w;\zp)V(\zp)G(\zp,\w;\zo).\label{LSEq}
\eeq
One method of solving this equation results in the generation of the Born series or, at least, the generation of Born approximations 
of various orders.
For instance, the first-order Born approximation (also known simply as \emph{the}
Born approximation) is obtained by approximating $G$ by $G_0$ above on the right only, yielding
\beq
G(z,\w;\zo)\approx G_0(z,\w;\zo)+\frac{\w^2}{c^2_0}\fint d\zp G_0(z,\w;\zp)V(\zp)G_0(\zp,\w;\zo).
\eeq
The second-order Born approximation can be obtained by first replacing $G$ on the right-hand side of eq.~(\ref{LSEq}) with the entirety of
the right-hand side, 
\bea
G(z,\w;\zo)\!\!\!\!\!&=&\!\!\!\!\!G_0(z,\w;\zo)+\frac{\w^2}{c^2_0}\fint\!\! d\zp G_0(z,\w;\zp)V(\zp)G_0(\zp,\w;\zo)\nonumber\\
& &\hspace{15mm}
+\frac{\w^4}{c^4_0}\fint\!\! d\zp \fint\!\! d\zpp G_0(z,\w;\zp)V(\zp)G_0(\zp,\w;\zpp)V(\zpp)G(\zpp,\w;\zo),\label{OnceIter}
\eea
and then setting $G\approx G_0$ on the right again.  The general rules for obtaining approximations of arbitrary order are
\bea
G^{(1)}(z,\w;\zo)&=&G_0(z,\w;\zo)\nonumber\\
G^{(n)}(z,\w;\zo)&=&G_0(z,\w;\zo)+\frac{\w^2}{c^2_0}\fint d\zp G_0(z,\w;\zp)V(\zp)G^{(n-1)}(\zp,\w;\zo),\label{BornIter}
\eea
where $G^{(n)}$ indicates the $n$-th approximant to $G$.
For our purposes, however, the once-iterated equation \emph{without} the substitution of $G_0$ for $G$, eq.~(\ref{OnceIter}), is what we
want.

To greatly shorten some of the equations to follow, we define some helpful notation.  Setting $z=z_0=0$, we let
\beq
\GVG=\frac{\w^2}{c^2_0}\fint d\zp G_0(0,\w;\zp)V(\zp)G_0(\zp,\w;0)
\eeq
and
\bea
\GVGVG&=&\frac{\w^4}{c^4_0}\fint\!\!\! d\zp \fint\!\!\! d\zpp G_0(0,\w;\zp)V(\zp)G_0(\zp,\w;\zpp)V(\zpp)G(\zpp,\w;0),\label{GVGVG}
\eea
where the subscript $RS$ indicates that the quantities are evaluated only for the given source/rec\-eiver geometry.  Thus, our 
once-iterated equations becomes
\beq
G(0,\w;0)=G_0(0,\w;0)+\GVG+\GVGVG.\label{OnceItOp}
\eeq
Using the explicit form of the Green's function $G_0$, 
\beq
G_0(z,\omega;z_0)=\frac{ic_0}{2\omega}e^{i\omega|z-z_0|/c_0}\label{G0},
\eeq
and the fact that $V(z)\equiv 0$ for all $z<0$, we find
\bea
\GVG&=&\frac{\omega^2}{c_0^2}\int_0^\infty dz \left(\frac{ic_0}{2\omega}\right)^2 V(z) e^{2i\omega|z|/c_0}\nonumber\\
&=&-\frac{1}{4}\sqrt{2\pi}\tilde{V}(2\omega/c_0),\label{GVGeq}
\eea
where $\tilde{V}$ is the Fourier transform of $V$.  Inverting this relationship, we find
\beq
V(z)=-\frac{4}{\pi c_0}\int_{-\infty}^{\infty} d\omega e^{-2i\omega z /c_0} \GVG.\label{Veq}
\eeq
(Note that our Fourier transform convention is that of symmetric normalization for the inverse and forward transforms.)
Letting $\Dw=G(0,\w;0)-G_0(0,\w;0)$ and
\beq
U(z)=-\frac{4}{\pi c_0}\int_{-\infty}^{\infty} d\omega e^{-2i\omega z /c_0}\Dw,\label{U}
\eeq
which is tantamount to a constant-velocity depth migration of the data, we apply the above inverse Fourier transform to the 
once-iterated equation and rearrange to get
\bea
V(z)=U(z)+\frac{4}{\pi c_0}\int_{-\infty}^{\infty} d\omega e^{-2i\omega z /c_0}\GVGVG.\label{FredEq}
\eea
Noting that $G$ on the right-hand side depends on $V$, we see that this equation is an inhomogeneous, {\em highly} non-linear
Fredholm integral equation of the second kind for the scattering potential $V$.  This equation is a novel result and forms the
basis for our approach to waveform inversion.

\subsection{An iterative method of solution}

There is no pre-existing mathematical apparatus for solving equations like eq.~(\ref{FredEq}).  One possible scheme for solving 
it---one that should eventually be vetted by mathematicians---can be obtained by analogy to the iterative way of solving 
the Lippmann-Schwinger equation that we discussed earlier (eqs.~\ref{BornIter}).  We propose to construct a sequence of approximants to $V$ 
by the following rules:
\bea
V^{(0)}(z)&\equiv& 0\\
V^{(n)}(z)&=&U(z)+\frac{4}{\pi c_0}\int_{-\infty}^{\infty}
d\omega e^{-2i\omega z /c_0}\left(\Go\V^{(n-1)}\Go\V^{(n-1)}\G^{(n-1)}\right)_{RS},
\eea
where we generalize the notation of eq.~(\ref{GVGVG}) by substituting $V^{(n-1)}$ for $V$ and use $G^{(n-1)}$ to indicate
the Green's function for a medium with potential $V^{(n-1)}(z)$.
Going back to the once-iterated Lippmann-Schwinger equation, eq.~(\ref{OnceItOp}), which holds for any medium, we have
\beq
\left(\Go\V^{(j)}\Go\V^{(j)}\G^{(j)}\right)_{RS}=G^{(j)}(0,\w;0)-G_0(0,\w;0)-\left(\Go\V^{(j)}\Go\right)_{RS}.
\eeq
Using this relation in our iterative equation, we can achieve a substantial simplification to our second iteration rule above:
\beq
V^{(n)}(z)=U(z)+V^{(n-1)}(z)+\frac{4}{\pi c_0}\int_{-\infty}^{\infty}d\omega e^{-2i\omega z /c_0}\left(G^{(n-1)}(0,\w;0)-G_0(0,\w;0)\right).
\eeq
In parallel with our definition of $U$, eq.~(\ref{U}), we set
\beq
U^{(n-1)}(z)=-\frac{4}{\pi c_0}\int_{-\infty}^{\infty} d\omega e^{-2i\omega z /c_0}\left(G^{(n-1)}(0,\w;0)-G_0(0,\w;0)\right)
\eeq
to get the final version of our iteration rules:
\bea
V^{(0)}(z)&\equiv& 0\nonumber\\
V^{(n)}(z)&=&U(z)+V^{(n-1)}(z)-U^{(n-1)}(z).\label{Vit}
\eea

The algorithmic interpretation of these iteration rules is simple.  At each step, the new approximant $V^{(n)}$ is obtained from the
previous one, the ``migrated'' data $U$, and the migrated synthetic data from a medium with scattering potential $V^{(n-1)}$.  Thus our
algorithm should recover $V$ through a deterministic sequence of forward modelling simulations.  It is for this reason that we call our 
approach to inverse scattering ``iterative inverse propagation''.

\section{Numerical Examples}

As proof of concept, we present three synthetic tests of 1D iterative inverse propagation.  In each case, we constructed a test model and
generated synthetic seismograms with a staggered-grid finite-difference (FD) code (see e.g. Virieux, 1986) using a Gaussian source wavelet 
with a half-width of about 0.5 s.  In the inversion, we used an initial constant background velocity of 1 km/s, and the necessary forward 
modelling was done with the same FD code and source wavelet that was used to generate the synthetic data.  We emphasize that {\em no} 
pre-processing, such as removal of multiples, was performed on the data.

\begin{figure}[htb]
\begin{center}
\epsfig{file=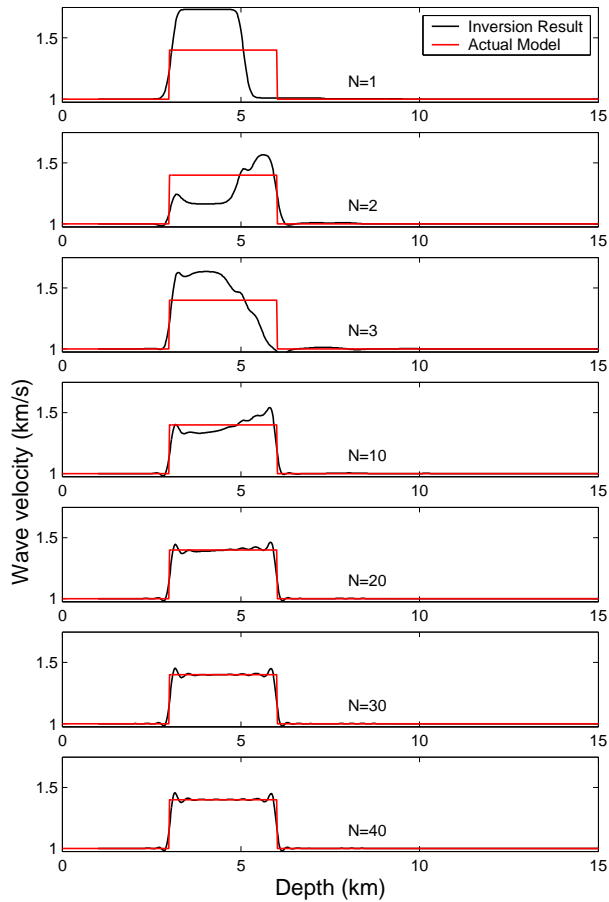,width=230pt,clip=}
\end{center}
\caption{One layer of thickness 3 km with a wave velocity $40\%$ above background.  Comparison of the true model (in red) and the 
inversion results (in black) are shown for various iterations indicated in each panel by the parameter $N$.}
\end{figure}

\begin{figure}[htb]
\begin{center}
\epsfig{file=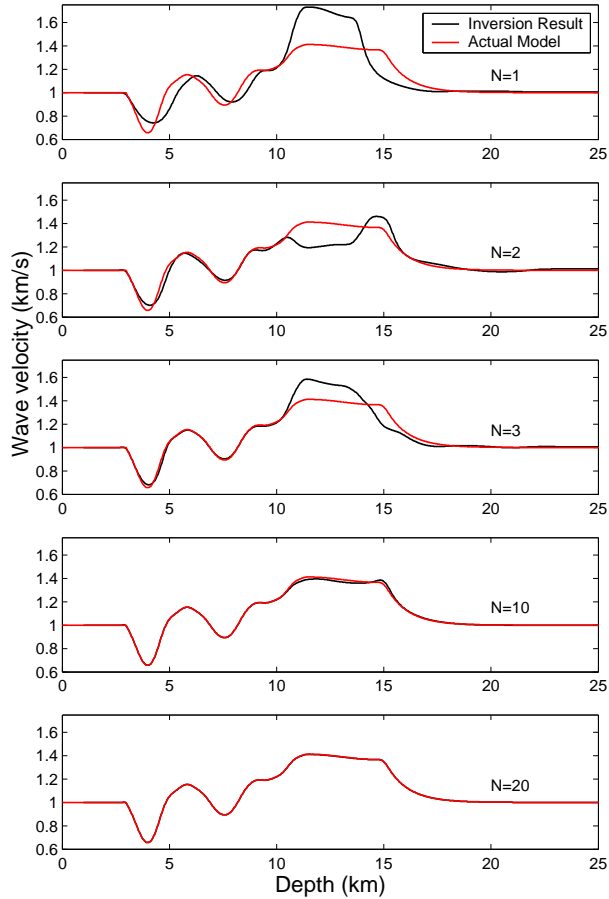,width=230pt,clip=}
\end{center}
\caption{A smoothed random model with velocity variations of approximately $\pm 40\%$ about background.  Note that the inversion 
recovers even the gradient at the end of the model.}
\end{figure}

\begin{figure}[htb]
\begin{center}
\epsfig{file=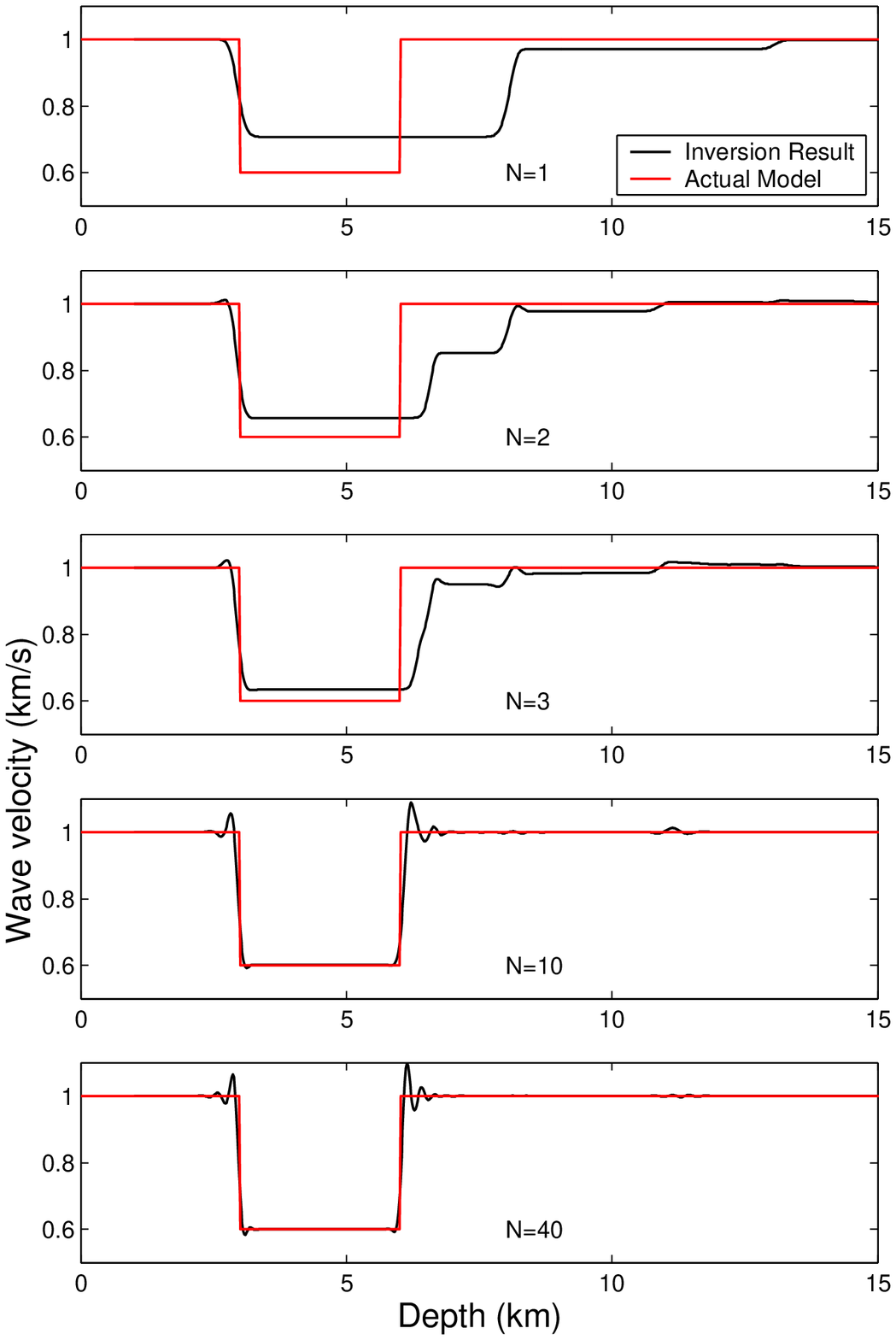,width=230pt,clip=}
\end{center}
\caption{One low-velocity layer of thickness 3 km with a wave velocity $40\%$ below background.  Note the progressive collapse 
of multiples into the main structure.}
\end{figure}

For numerical reasons, one deviation from the derived iteration rules, eq.~(\ref{Vit}), was used.  At each iteration,
the model obtained from the previous step was used as the new ``mapping velocity'', i.e., the velocity used to map time into space.  In other
words, instead of using the factor of $2z/c_0$ in the exponential of eq.~(\ref{Vit}), a more general function $T(z)$---the traveltime
from the source to the point $z$ and back to the receiver---was computed from the previous iteration.  This very practical {\em ad hoc}
feature was used to eliminate some instabilities that were otherwise occurring at the bottom end of the structure of any test attempted.  
In fact, this approach also accelerated convergence and indicates the need to incorporate non-constant reference models in future work.

Figure 1 shows the results for our first test, a model with a single high-velocity layer imbedded in an otherwise homogeneous medium.  
The results of the first three iterations and iterations 10, 20, 30, and 40 are shown.  Both the amplitude of the velocity jump ($40 \%$ above 
reference) and the positions of the top and bottom of the layer are quite well recovered, with the essential features of the true model 
recovered by iteration 20 and only moderate improvement obtained thereafter.  Those deviations from the true model that are present are due 
to the band-limited wavelet used in both the initial synthetic data and the FD modelling performed in the inversion.

Figure 2 demonstrates the ability of the algorithm to reconstruct the details of a complicated, smooth model with fluctuations of approximately
$\pm 40\%$ about the reference velocity.  We see that by iteration 10 the errors in the inversion result are negligible.  It is
particularly notable that even the smooth gradient at the end of the model is recovered.

Finally, Figure 3 shows the results for a single low-velocity layer of velocity $40 \%$ below reference.  The value of this example is that it
shows clearly what happens to multiples as the inversion progresses.  Specifically, one sees that they are gradually ``pushed'' upward into
the main structure, indicating that the information they provide is instrumental in obtaining an accurate final result.

\section{Discussion}

We have shown the potential of iterative inverse propagation to recover the details of wave velocity structures in 1D.  There are, of course,
going to be many hurdles to overcome in extending this method to practical use.  In addition to incorporating non-constant reference media,
we must also accommodate irregular source/receiver geometries, elasticity, attenuation, limited frequency ranges, and unknown source wavelets,
just to name a few.  There will also be the issue of the enormous computational effort necessary to perform the forward wave propagation
computations required by the algorithm.  

However, it is likely that all of these will be overcome in time, and the straightforward nature of a deterministic algorithm such as this
implies substantial long-term benefits to seismic exploration.  

\section*{Acknowledgments}

I wish to thank Jacques Leveille, Scott Morton, and Gene Humphreys for helpful discussions.  I also wish to thank Ru-Shan Wu for his support
and encouragement in the latter stages of this work.



\begin{thebibliography}{}

\bibitem{1} Carvalho, P. M., 1992, Free Surface Multiple Reflection Elimination Method Based on Nonlinear Inversion of Seismic Data: 
Ph.D. thesis, Universidad Federal de Bahia.

\bibitem{2} Jost, R., and W. Kohn, 1952, Construction of a potential from a phaseshift: Physical Review, {\bf 87}, 977--992.

\bibitem{3} Moses, H. E., 1956, The calculation of the scattering potential from reflection coefficients: Physical Review, {\bf 102},
559--567.

\bibitem{4} Pratt, R. G., 1999a, Seismic waveform inversion in the frequency domain, Part 1: Theory and verification
in a physical scale model: Geophysics, {\bf 64}, 888--901.

\bibitem{5} Pratt, R. G., 1999b, Seismic waveform inversion in the frequency domain, Part 2: Fault delineation in
sediments using crosshole data: Geophysics, {\bf 64}, 901--913.

\bibitem{6} Prosser, R. T., 1980, The formal solutions of inverse scattering problems III: Journal of Mathematical Physics, {\bf 21},
2648--2653.

\bibitem{7} Razavy, M., 1975, Determination of the wave velocity in an inhomogeneous medium from reflection data:
Journal of the Acoustical Society of America, {\bf 58}, 956--963.

\bibitem{8} Sen, M. K., and P. L. Stoffa, 1995, Global Optimization Methods in Geophysical Inversion: Elsevier
Science Publications.

\bibitem{9} Sirgue, L., and R. G. Pratt, 2004, Efficient waveform inversion and imaging: A strategy for
selecting temporal frequencies: Geophysics, {\bf 69}, 231--248.

\bibitem{10} Virieux, J., 1986, P-SV wave propagation in heterogeneous media:~Velo\-city-stress finite-difference
method: Geophysics, {\bf 51}, 889--901.

\bibitem{11} Weglein, A. B., F. V. Araujo, P. M. Carvalho, R. H. Stolt, K. H. Matson, R. Coates, D. Corrigan, D. J. Foster, S. A. Shaw, and H.
Zhang, 2003, Inverse scattering series and seismic exploration: Inverse Problems, {\bf 19}, R27--R83.

\bibitem{12} Yokota, T., and J. Matsushima, 2004, Seismic waveform tomography in the frequency-space
domain: selection of the optimal temporal frequency for inversion: Exploration Geophysics, {\bf 35}, 19--24.

\end{thebibliography}
\end{document}